\documentclass[12pt]{article}
\usepackage{epsfig}

\newcommand{\MeV}{{\,\rm MeV\/}}
\newcommand{\GeV}{{\,\rm GeV\/}}
\newcommand{\slD}{\kern1pt/\kern-8pt D}

\newcommand{\slP}{\kern2pt\raise1.5pt\hbox{/}\kern-8pt P}
\newcommand{\slk}{/\kern-6pt k}
\newcommand{\sll}{/\kern-4pt l}
\newcommand{\slp}{p\kern-5pt/}
\newcommand{\slq}{q\kern-5.5pt/}
\newcommand{\slv}{v\kern-5pt\raise1pt\hbox{$\scriptstyle/$}\kern1pt}

\newcommand{\pfrac}[2]{\left(\frac{#1}{#2}\right)}

\begin{document}
\begin{flushright}
MZ-TH/00-36, UM-TH-00-40, hep-ph/0008247
\end{flushright}
\begin{center}
{\Large\bf A dynamic mass definition\\[3pt]
  near the $t\bar t$ threshold}\\[12pt]
{\large Stefan Groote$^a$ and Oleg Yakovlev$^b$}\\[7pt]
$^a$ Institut f\"ur Physik der Johannes-Gutenberg-Universit\"at,\\[3pt]
Staudinger Weg 7, 55099 Mainz, Germany\\[7pt]
$^b$ Randall Laboratory of Physics, University of Michigan,\\[3pt]
  Ann Arbor, Michigan 48109-1120, USA
\end{center}
\tableofcontents
\begin{abstract}
Starting from general considerations about the different concepts which exist
for the definition of the quark mass near the pair production threshold, we
introduce the definition of a mass with a dynamic degree of freedom that
extends the previously given concept of the static PS mass.
\end{abstract}

\vfill\noindent
{\em Talk given by Stefan Groote at the Laboratory of Nuclear Studies,\\
  Cornell University, Ithaca, NY, August 23rd, 2000}
\newpage

\section{Keeping up with experiments}
Besides existing machines and a whole bunch of data sets, further lepton pair
colliders such as the Next Linear Collider (NLC) and the Future Muon Collider
(FMC) are designed for the future. They will be able to determine the
properties of heavy particles to an extend never reached by present colliders.
Especially they will be able to measure the properties of the top quark which
was first measured at the Tevatron with a mass of $m_t=174.3\pm 5\GeV$
(see e.g.~\cite{Tevatron} and references therein). Although the top quark
will be studied at the LHC and the Tevatron (RUN-II) as well with an expected
accuracy for the mass of $2-3\GeV$, the measurements at NLC are expected to
have an accuracy of $0.1\%$ ($100-200\MeV$)~\cite{Peskin}.

Theorists are therefore asked to improve their predictions, too. Of special
interest is the consideration of the cross section near the pair production
threshold because the resonance observed there is constituted solely by a
colour singlet state. A lot has been done in the past and just recently to
calculate and match corresponding diagrams up to the next-to-next-to-leading
order in order to determine this cross section (for an overview on work that
has been done see e.g.\ Ref.~\cite{YG}). But the question remains what we
actually want to determine. What can we say about the mass of a particle which
never will show up in nature in isolated fashion because of its confinement?
Does the term of {\em the\/} mass of the top quark really makes sense?

The everydays life of theorists show that this actually makes no sense. There
are different concepts on the market -- each of them with different advantages
and disadvantages, regions of validy and failure. They all show that there
cannot be a unique definition of the quark mass. Instead the task of the
theory can only be to quantify and limit the theoretical uncertainties to an
amount comparable to the experimental measurements. One of these theory-born
uncertainties are special kinds of IR singularities known as renormalons which
limit the accuracy of a prediction to the order of $\Lambda_{\rm QCD}$. Just
recently this problem has been overcome (a synapsis of the progress is given
in Ref.~\cite{Review}), and this talk will deal with the different concepts
which were developed in this context, closing with an own suggestion
presented by the authors~\cite{YG}.

\section{$\overline{\rm MS}$ mass and on-shell or pole mass}
A mass quite often used in perturbation theory is the $\overline{\rm MS}$ mass.
It is related to the bare mass of the underlying QCD by the calculation of self
energy diagrams of the quark. The term $\overline{\rm MS}$ mass indicates that
for this calculation we use the $\overline{\rm MS}$ subtraction scheme which
itself is based on the dimensional regularization method. The calculation of
self energy diagrams which is done actually up to the third order QCD
perturbation theory~\cite{ChetyrkinSteinhauser,MelnikovRitbergen} leads to a
couple of master diagrams which are calculated at center-of-mass energies in
the deep Euklidean domain ($q^2<0$, $|q|$ large) where the perturbation
series converge sufficiently good. This $\overline{\rm MS}$ mass of course
does not fit the range around the threshold and spoils the non-relativistic
expansion if applied. But the $\overline{\rm MS}$ mass can be continued to
this region.

The continuation is done by the Borel transformation. This transform can be
understood as the extrapolation from the deep Euklidean domain to the
$t\bar t$ threshold domain where a simultaneous limit of the negative
center-of-mass energy and the order of the used derivative is taken. The
ratio of these two quantities remains fixed and constitutes the Borel
parameter. As shown in Ref.~\cite{AgliettiLigeti}, this Borel transformation
can be made explicit for the case of the ``large $\beta_0$ limit'' where the
prescription of Brodsky, Lepage and Mackenzie~\cite{BrodskyLepageMackenzie} is
used to sum up all fermion loop insertions (being proportional to the zeroth
order QCD beta function coefficient $\beta_0=11C_A/3-2N_F/3$) in order to
obtain an estimate (for a relation of this estimate to the $\overline{\rm MS}$
mass see Ref.~\cite{BenekeBraun}).

If we use the prescription for the static quark-antiquark potential $V(r)$,
we obtain
\begin{equation}
V(r)=\sum_{n=0}^\infty V_n(\beta_0\alpha_s)^{n+1}.
\end{equation}
The Borel transformation of this series expression leads to
\begin{equation}
\tilde V(r;u)=\sum_{n=0}^\infty\frac{V_n}{n!}u^n
\end{equation}
The series for the Borel transformed quantity converges even if the
perturbative series for $V(r)$ itself is divergent. The inverse Borel
transformation
\begin{equation}
V(r)=\int_0^\infty e^{u/(\beta_0\alpha_s)}\tilde V(r;u)du
\end{equation}
provides then a good definition. However, there might occur singularities on
the integration contour. As mentioned before, for the ``large $\beta_0$''
estimate (also known as ``naive non-abelianization'') we obtain an explicit
expression
\begin{equation}
\tilde V(r;u)\sim\frac{\Gamma(1/2+u)\Gamma(1/2-u)}{\Gamma(2u+1)}
\end{equation}
which has simple poles for $u=k+1/2$, $k=0,1,2,\ldots$. These singularities
are a special case of IR renormalons.

The same happens then for the continuation of the $\overline{\rm MS}$ mass to
the on-shell mass in the threshold region. This on-shell mass is appropriate
to the treatment near the threshold because the quark is then assumed to be
nearly on-shell. The on-shell mass $m$ (or pole mass as it is called more
often) can be determined by the demand that the inverse Fermi propagator
should vanish at this point,
\begin{equation}
(S_F(q))^{-1}\Big|_{q^2=m^2}=0.
\end{equation}
But the continuation of the $\overline{\rm MS}$ mass via Borel transformation 
to the pole mass leads again to IR renormalon ambiguities.

So there is no way out? Wait, there is one! Both the pole mass and the static
potential show an IR renormalon ambiguity. Maybe they cancel in a specific
combination of these two quantities? Indeed it emerges that in the combination
$2m+V(r)$ the IR sensitivity vanishes. Only for this reason it is then
possible to construct a quantity which has no such ambiguities. In fact there
are several of them, called threshold masses which we will deal with in the
following.

\section{Threshold mass definitions}
The first definition we want to mention here is the definition given by
Hoang and Teubner~\cite{HoangTeubner}. We mention it first because it is the
most sophisticated definition. Starting with the non-relativistic QCD
(NRQCD)~\cite{Lepage}, an effective theory designed to handle the
non-relativistic quark-antiquark system which separates the low-momentum
scales $mv$ and $mv^2$ from the high momentum scale $m$ ($v$ is the velocity
of the quark), they stress the relevance of the different scales $v$ and
$\alpha_s$ and their interrelation. Especially in the non-relativistic regime
even the space and time components can be scaled differently. So there are
``soft'' ($q\sim m(v;v)$), ``potential'' ($q\sim m(v^2;v)$), and ``ultrasoft''
regions ($q\sim m(v^2;v^2)$) where the last one is populated only by gluons,
not by quarks. If we separate the dynamical degrees of freedom also from the
``soft'' heavy quarks and gluons and the ``potential'' gluons, supplemented
by an expansion in momentum components to the order $mv^2$, this leads to
the construction of the ``potential NRQCD'' (PNRQCD)~\cite{PinedaSoto} with
the Langrangian
\begin{equation}
{\cal L}_{\rm PNRQCD}=\tilde{\cal L}_{\rm NRQCD}+\int(\psi^\dagger\psi)
  (\vec r\,)V(\vec r\,)(\chi^\dagger\chi)(\vec 0)d^3r.
\end{equation}
Here the heavy quark-antiquark potential $V(\vec r\,)$ occurs, $\psi$ and
$\chi$ are the quark and antiquark spinors, respectively. Finally the authors
of Ref.~\cite{HoangTeubner} come to the conclusion that an adequate threshold
mass definition is given by one half of the perturbative mass of a fictious
toponium $1^3S_1$ ground state. This threshold mass definition is referred to
in the literature as $1S$ mass.

A second threshold mass definition is given by Bigi {\it et al.}~\cite{Bigi}
and is known by the name ``low scale'' (LS) mass.

Finally, Beneke gave a definition which then led to the generalization
suggested by us. The static potential subtracted (PS) mass~\cite{Beneke} is
the minimal version of a threshold mass. As the name tells us, the definition
is given by combining IR singular parts taken from the static quark-antiquark
potential with the pole mass in order to cancel out the IR sensitivity.
Starting point is the observation that the IR sensitive parts of the potential
stem from long distance contributions, i.e.\ small values of the
three-momentum. If we therefore restrict the three-momentum range of the
potential as expressed by the Fourier transformation 
\begin{equation}
V(|\vec r\,|)=\int\frac{d^3q}{(2\pi)^3}e^{i\vec q\cdot\vec r}
  \tilde V(|\vec q\,|)
\end{equation}
by $|\vec q\,|<\Lambda_{\rm QCD}$, this results (for the leading order
Coulomb potential) in the contribution
\begin{equation}
\int_{|\vec q\,|<\Lambda_{\rm QCD}}\frac{d^3q}{(2\pi)^3}e^{i\vec q\cdot\vec r}
  \left(-\frac{4\pi\alpha_sC_F}{\vec q\,^2}\right)
  =-\frac{2\alpha_sC_F}\pi\Lambda_{\rm QCD}+O(\Lambda_{\rm QCD}^3r^2)
\end{equation}
which restricts the accuracy. The way out is to use a potential $V(r;\mu_f)$
defined by the Fourier transformation with $|\vec q\,|>\mu_f$ and to subsume
the remaining part to the mass,
\begin{equation}
m_{\rm PS}(\mu_f)=m-\delta m_{\rm PS}(\mu_f),\qquad
\delta m_{\rm PS}(\mu_f)=-\frac12\int_{|\vec q\,|<\mu_f}\frac{d^3q}{(2\pi)^3}
  \tilde V(|\vec q\,|)
\end{equation}
(it is legitimate to use $e^{i\vec q\cdot\vec r}\approx 1$ here). The scale
$\mu_f$ introduced by this is called {\em factorization scale\/}. It drops out
again if the static PS mass is related to the $\overline{\rm MS}$ mass.

\section{A dynamic mass definition}
It has been pointed out already in Ref.~\cite{Beneke} that the replacement
\begin{equation}
\frac1{v\cdot q}\rightarrow -i\pi\delta(v\cdot q)
\end{equation}
in the self energy leads to the leading infrared behaviour. We have taken this
consideration as starting point for our definition and extended it to the more
general non-static case.

\begin{figure}
\begin{center}
\epsfig{figure=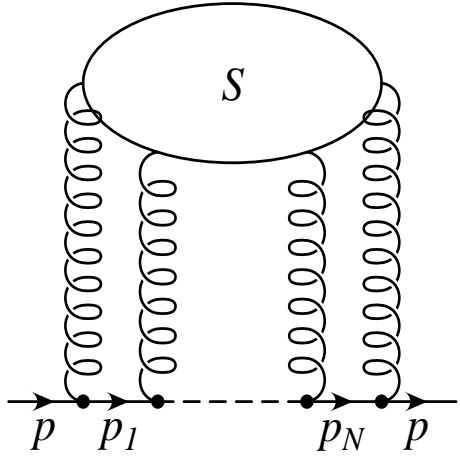, height=6truecm,width=6truecm}
\caption{\label{fig1}
  The general structure of the self energy diagram of a quark} 
\end{center}
\end{figure}

The general structure of the self energy diagram for an on-shell quark with
mass $m$ and momentum $p$ ($p^2=m^2$) is shown in Fig.~\ref{fig1}. The
expression for this diagram is given by
\begin{eqnarray}\label{def1}
-i\Sigma(\slp)&=&\int\prod_{m=1}^M\frac{d^4l_m}{(2\pi)^4}
  S^{\{a_n\}}_{\{\alpha_n\}}(\{l_m\})\left(-ig_s\gamma^{\alpha_{N+1}}
  T_{a_{N+1}}\right)\ \times\nonumber\\&&\qquad\qquad\qquad\times\
  \prod_{n=N}^1\frac{i}{\slp_n-m}\left(-ig_s\gamma^{\alpha_n}T_{a_n}\right)
\end{eqnarray}
where the last factor is a non-commutative product with decreasing index $n$.
The line momenta $k_n$ are linear combinations of the gluon loop momenta $l_m$,
the particular representation is specified by the structure $S$. The symbol
$\{l_m\}$ means the set of all these loop momenta, the same symbol is used
for the Lorentz and colour indices. In general we have $M\le N$ which means
that line momenta can be correlated. The momenta of the virtual quark states
are given by $p_n=p+k_n$.

The quark is considered to be at rest, $p=(m,\vec 0)$, and it interacts with a
number of gluons. The subdiagram $S$ displayed in Fig.~\ref{fig1} describes
the interaction between the gluons. In general the quark lines between the
interaction points represent virtual quark states. However, if the virtual
quark comes very close to the mass shell and the total momentum of the cloud
of virtual gluons becomes soft, this situation gives rise to long-distance
nonperturbative QCD interactions. The described (virtual) contributions
results in the soft part of the self energy, $\Sigma_{\rm soft}$. So we define
\begin{eqnarray}\label{def2}
\lefteqn{-i\Sigma_{\rm soft}(\slp)\ =\ \sum_{i=1}^N\int\prod_{m=1}^M
  \frac{d^4l_m}{(2\pi)^4}S^{\{a_n\}}_{\{\alpha_n\}}(\{l_m\})
  \left(-ig_s\gamma^{\alpha_{N+1}}T_{a_{N+1}}\right)\
  \times}\nonumber\\&&\times\ \prod_{n=N}^{i+1}
  \frac{i}{\slp_n-m}\left(-ig_s\gamma^{\alpha_n}T_{a_n}\right)
  i(\slp_i+m)\left(-i\pi\delta(p_i^2-m^2)\right)\
  \times\nonumber\\&&\qquad\times\
  \left(-ig_s\gamma^{\alpha_i}T_{a_i}\right)\prod_{n=i-1}^1
  \frac{i}{\slp_n-m}\left(-ig_s\gamma^{\alpha_n}T_{a_n}\right).
\end{eqnarray}
One can derive this expression from Eq.~(\ref{def1}) by using the identity
\begin{equation}\label{ident1}
\frac1{p^2-m^2+i\epsilon}=-i\pi\delta(p^2-m^2)+P\pfrac1{p^2-m^2}
\end{equation}
and the fact that the principal value integral does not give any infrared
sensitive contribution. The delta function can be used to remove the
integration over the zero component of $k_i$. In order to parameterize the
softness of the gluon cloud we impose a cutoff on the spatial component,
$|\vec k_i|<\mu_f$, and indicate this by the label $\mu_f$ for the
factorization scale written at the upper limit of the three-dimensional
integral. So we can rewrite Eq.~(\ref{def2}) as
\begin{equation}
\Sigma_{\rm soft}(\slp,\mu_f)=-\frac12\sum_{i=1}^N\int^{\mu_f}
  \frac{d^3k_i}{(2\pi)^3}V(\vec k_i,p)
\end{equation}
where
\begin{eqnarray}
\lefteqn{V(\vec k_i,p)\ =\ -\int\prod_{m=1}^{M-1}\frac{d^4l_m}{(2\pi)^4}
  S^{\{a_n\}}_{\{\alpha_n\}}(\{l_m\})
  \left(-ig_s\gamma^{\alpha_{N+1}}T_{a_{N+1}}\right)\
  \times}\\&&\kern-24pt\times\ \prod_{n=N}^{i+1}
  \frac{i}{\slp_n-m}\left(-ig_s\gamma^{\alpha_n}T_{a_n}\right)
  \frac{\slp_i+m}{2p_i^0}
  \left(-ig_s\gamma^{\alpha_i}T_{a_i}\right)\prod_{n=i-1}^1
  \frac{i}{\slp_n-m}\left(-ig_s\gamma^{\alpha_n}T_{a_n}\right)\nonumber
\end{eqnarray}
and finally define
\begin{equation}
m_{\overline{\rm PS}}=m-\delta m_{\overline{\rm PS}}\quad\mbox{with}\quad
  \delta m_{\overline{\rm PS}}=\Sigma_{\rm soft}(\slp)\Big|_{\slp=m}.
\end{equation}
In the following we deal with the different realizations of this compact
expression. As we will see explicitly, the function $V(\vec k,p)$ occurring as
integrand can be seen as quark-antiquark potential where we have summed over
the spin of the tensor product of a final state with an initial state. Because
the static quark-antiquark potential is used in a similar way in
Ref.~\cite{Beneke}, we recover the result of Ref.~\cite{Beneke} in the static
limit.

\section{The one-loop contribution}
\begin{figure}
\begin{center}
\epsfig{figure=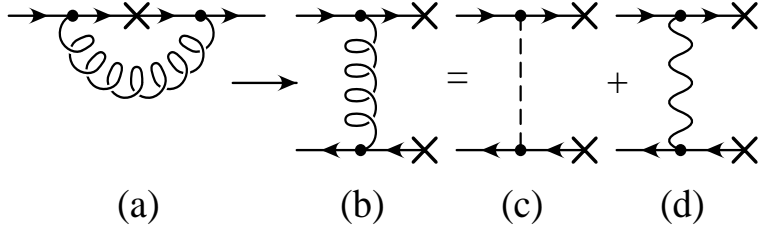, height=3truecm,width=10truecm}
\caption{\label{fig2}Leading order contribution to the quark self energy
(a) and to the quark-antiquark potential (b). The cross indicates the
point where we cut the quark line by imposing an on-shell condition to
the virtual quark state. In Coulomb gauge the gluon propagator can be
decomposed in a Coulomb propagator (c) and a transverse propagator (d).}
\end{center}
\end{figure}

The leading order contribution to the self energy of the quark is given by
\begin{equation}
\Sigma(\slp)=i\int\frac{d^4k}{(2\pi)^4}(-ig_s\gamma_\alpha T_a)
  \frac{i}{\slp+\slk-m}(-ig_s\gamma^\alpha T_a)\frac{-i}{k^2}
\end{equation}
where Feynman gauge is used for the gluon. The soft part of it is given by
\begin{equation}\label{leadsoft}
\Sigma_{\rm soft}(\slp,\mu_f)=-\frac12\int^{\mu_f}\frac{d^3k}{(2\pi)^3}
  V(\vec k,p),\qquad V(\vec k,p)=V_+(\vec k,p)+V_-(\vec k,p)
\end{equation}
where
\begin{equation}\label{Vpm}
V_\pm(\vec k,p)=-\frac{g_s^2C_F(\sqrt{m^2+\kappa^2}\mp 2m)}{2m
  \sqrt{m^2+\kappa^2}(\sqrt{m^2+\kappa^2}-m)}
\end{equation}
and $\kappa=|\vec k|$. The procedure of taking the soft part by setting a cut
is shown in Fig.~\ref{fig2}(a--b).
The two potential parts are known as the scattering potential $V_+(\vec k,p)$
and the annihilation potential $V_-(\vec k,p)$ and correspond to the two
zeros $k_\pm=\pm\sqrt{\kappa^2+m^2}-m$ of the delta function. Integrating the
potential up to the factorization scale $\mu_f$, we obtain
\begin{equation}
\Sigma_{\rm soft}(\mu_f)=\frac{\alpha_sC_F}{2\pi} m
  \left\{3\ln\left(\frac{\mu_f}m+\sqrt{\frac{\mu_f^2}{m^2}+1}\,\right)
  -\frac{\mu_f}m\sqrt{\frac{\mu_f^2}{m^2}+1}\,\right\}.
\end{equation}
The expansion of this expression in small values of $\mu_f/m$ results in
\begin{equation}
\Sigma_{\rm soft}(\mu_f)=\frac{\alpha_sC_F}{\pi}\mu_f
  \left\{1-\frac{\mu_f^2}{2m^2}+O\pfrac{\mu_f^4}{m^4}\right\}.
\end{equation}
The first term reproduces the result given in Ref.~\cite{Beneke} to leading
order in $\alpha_s$ while the second term is the recoil correction to the
static limit in this order of perturbation theory. This second term is related
to the Breit-Fermi potential but does not coincide with it. 

\section{Two-loop contributions}
\begin{figure}
\begin{center}
\epsfig{figure=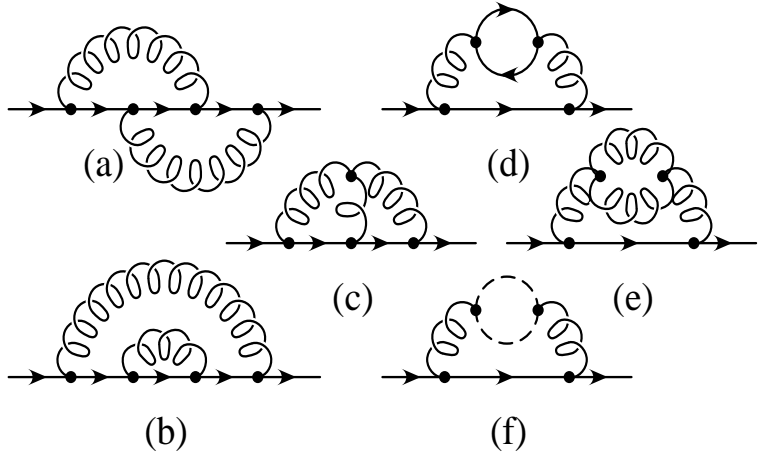, height=6truecm,width=10truecm}
\caption{\label{fig3}Two-loop contributions to the quark self energy}
\end{center}
\end{figure}

To take a step beyond the leading order perturbation theory, we consider
two-loop diagrams for the heavy quark self energy as shown in Fig.~\ref{fig3}.
In cutting the quark line in all possible ways we obtain a lot of diagrams
from the ones shown in Fig.~\ref{fig3}. However, we find that the final
contribution of the two abelian diagrams in Fig.~\ref{fig3}(a--b) to the
soft part of the self energy is suppressed by $\mu_f^2/m^2$. Therefore it
turns out that the only abelian diagrams which can give a non-suppressed
contribution to the soft part of the quark self energy are the diagrams
containing the vacuum polarization of the gluon as shown in
Fig.~\ref{fig3}(d--f). The simple calculation of these diagrams within the
$\overline{\rm MS}$ scheme, accounting only for light fermion loops, gluon
loop (and ghost loop if Feynman gauge is used) results after renormalization
in
\begin{eqnarray}
\Sigma_{\rm soft}^A&=&-\frac12\int^\mu\frac{d^3k}{(2\pi)^3}
  \left(-\frac{4\pi\alpha_s(\mu)C_F}{|\vec k|^2}\right)
  \left\{1+\frac{\alpha_s(\mu)}{4\pi}\left(a_1-\beta_0
  \ln\pfrac{|\vec k|^2}{\mu^2}\right)\right\}\nonumber\\
  &=&\frac{\alpha_s(\mu)C_F}\pi\mu_f\left\{1+\frac{\alpha_s(\mu)}{4\pi}
  \left(a_1-\beta_0\ln\pfrac{\mu_f^2}{\mu^2}\right)\right\}
\end{eqnarray}
with the same coefficient $a_1$ as in Ref.~\cite{Fischler} (see below). This
result has been anticipated because the expression in the curly brackets of
the integrand reproduces the next-to-leading order correction to the QCD
Coulomb potential.
 
For the only non-abelian diagram in Fig.~\ref{fig3}(c) we obtain
\begin{equation}
\Sigma_{\rm soft}^{NA}=\frac{\alpha_s^2C_FC_A}{8m}\mu_f^2.
\end{equation}
This result has been anticipated, too, to be minus one half of the non-abelian
correction to the QCD Coulomb potential, which is known in the literature 
(see for example Refs.~\cite{Gupta,Duncan}),
\begin{eqnarray}
\Sigma_{\rm soft}^{NA}&=&-\frac12\int^{\mu_f}\frac{d^3k}{(2\pi)^3}
  \left\{-\frac{\pi^2\alpha_s^2C_FC_A}{m|\vec k|}\right\}
  \ =\ \frac{\alpha_s^2C_FC_A}{8m}\mu_f^2.
\end{eqnarray}

\section{Our final result}
Summarizing all contribution up to NNLO accuracy, we obtain
\begin{eqnarray}\label{psmass}
m_{\overline{\rm PS}}(\mu_f)-m&=&-\frac{\alpha_s(\mu)C_F}\pi\mu_f
  \Bigg\{1+C'_0\frac{\mu_f}m+C''_0\frac{\mu_f^2}{m^2}\,+\nonumber\\&&\qquad
  +\frac{\alpha_s(\mu)}{4\pi}\left(C_1+C'_1\frac{\mu_f}m\right)
  +C_2\pfrac{\alpha_s(\mu)}{4\pi}^2\Bigg\}
\end{eqnarray}
(a result to a not yet specified order in $\mu_f/m$) where $m$ is the pole
mass, $\mu$ is the renormalization scale, $\mu_f$ is the factorization scale,
and
\begin{eqnarray}
C_0&=&1,\qquad C'_0\ =\ 0,\qquad C''_0\ =\ -\frac12,\nonumber\\
C_1&=&a_1-2\beta_0\ln\pfrac{\mu_f}\mu,\qquad
  C'_1\ =\ C_A\frac{\pi^2}2,\nonumber\\
C_2&=&a_2-2(2a_1\beta_0+\beta_1)\left(\ln\pfrac{\mu_f}\mu-1\right)
  \,+\nonumber\\&&\qquad
  +4\beta_0^2\left(\ln^2\pfrac{\mu_f}\mu-2\ln\pfrac{\mu_f}\mu+2\right).
\end{eqnarray}
The constants $a_1$ and $a_2$ are given by~\cite{Fischler,Peter,Schroder}
\begin{eqnarray}
a_1&=&\frac {31}9C_A-\frac{20}9T_FN_F,\nonumber\\
a_2&=&\left(\frac{4343}{162}+4\pi^2-\frac{\pi^4}4
  +\frac{22}3\zeta_3\right)C_A^2
  -\left(\frac{1798}{81}+\frac{56}3\zeta_3\right)C_AT_FN_F\,+\nonumber\\&&
  +\left(\frac{20}9T_FN_F\right)^2
  -\left(\frac{55}3-16\zeta_3\right)C_FT_FN_F,
\end{eqnarray}
the coefficients of the beta function are known as
\begin{eqnarray}
\beta_0=\frac{11}3C_A-\frac43T_FN_F,\quad
\beta_1=\frac{34}3C_A^2-\frac{20}3C_AT_FN_F-4C_FT_FN_F
\end{eqnarray}
where $C_F=4/3$, $C_A=3$ and $T_F=1/2$ are colour factors and $N_F=5$ is the
number of light flavours. The coefficients $C_1$ and $C_2$ in
Eq.~(\ref{psmass}) have been derived in Ref.~\cite{Beneke} by using known
corrections to the QCD potential. In Ref.~\cite{YG} we have obtained the
coefficients $C'_0$, $C''_0$, and $C'_1$. Note that our result can be
represented in a condensed form as
\begin{equation}
m_{\overline{\rm PS}}(\mu_f)-m=-\frac12\int^{\mu_f}\frac{d^3k}{(2\pi)^3}
  \left(V_C(|\vec k|)+ V_R(|\vec k|)+V_{NA}(|\vec k|)\right)
\end{equation}
where the first term $V_C$ is the static Coulomb potential, $V_R$ is the
relativistic correction (which is related to Breit-Fermi potential but does
not coincide with it), and $V_{NA}$ is the non-abelian correction. Note that
we did not include electroweak corrections in our calculations.

\begin{table}[t]
\begin{center}
\begin{tabular}{|c||c|c|c||c|c|c|}\hline&&&&&&\\
$\overline{m}(\overline{m})$&$m_{\rm PS}^{\rm LO}$&$m_{\rm PS}^{\rm NLO}$&
$\displaystyle\matrix{m_{\rm PS}^{\rm NNLO}\cr
  m_{\overline{\rm PS}}^{\rm NNLO}}$&$m_{\rm pole}^{\rm LO}$&
  $m_{\rm pole}^{\rm NLO}$&
  $m_{\rm pole}^{\rm NNLO}$\\&&&&&&\\\hline
\hline &&&&&&\\
  160&166.51&167.69&$\displaystyle\matrix{167.97\cr 167.95}$&
  167.44&169.05&169.56\\
\hline &&&&&&\\
  165&171.72&172.93&$\displaystyle\matrix{173.22\cr 173.20}$&
  172.64&174.28&174.80\\
\hline &&&&&&\\
  170&176.92&178.17&$\displaystyle\matrix{178.47\cr 178.45}$&
  177.84&179.52&180.05\\
\hline
\end{tabular}
\caption{\label{tab1}Top quark mass relations for the $\overline{\rm MS}$, PS,
$\overline{\rm PS}$, and the pole mass at LO, NLO, and NNLO  in $\GeV$. We fix
the $\overline{\rm MS}$ mass to be $\overline{m}(\overline{m})=160\GeV$,
$165\GeV$, and $170\GeV$ and find the pole mass at LO, NLO, and NNLO from the
three-loop relation in Refs.~\cite{ChetyrkinSteinhauser,MelnikovRitbergen}.
The PS and $\overline{\rm PS}$ masses are derived from the pole mass by using
Eq.~(\ref{psmass}) (without or with the $1/m$ contributions, respectively). We
use the QCD coupling constant $\alpha_s(m_Z)=0.119$~\cite{PDG},
$\mu=\overline{m}(\overline{m})$, and $\mu_f=20\GeV$.}
\end{center}
\end{table}

Using the relation given by Eq.~(\ref{psmass}) between $m_{\overline{\rm PS}}$
and $m$ as well as the relation between $m$ and $\overline{m}(\overline{m})$
given in Refs.~\cite{ChetyrkinSteinhauser,MelnikovRitbergen}, we fix the
$\overline{\rm MS}$ mass to take the values
$\overline{m}(\overline{m})=160\GeV$, $165\GeV$, and $170\GeV$ and determine
the pole mass at LO, NLO, and NNLO. This pole mass is then used as input
parameter $m$ in Eq.~(\ref{psmass}) to determine the PS and
$\overline{\rm PS}$ masses at LO, NLO, and NNLO. The results of these
calculations are collected in Table~\ref{tab1}. Taking the $\overline{\rm PS}$
mass instead of the PS mass, we observe an improvement of the convergence. The
differences for the mass values in going from LO to NLO to NNLO for e.g.\
$\overline{m}(\overline{m})=165\GeV$ read $7.64\GeV$, $1.64\GeV$, and
$0.52\GeV$ for the pole mass, $6.72\GeV$, $1.21\GeV$, and $0.29\GeV$ for the
PS mass and finally $6.72\GeV$, $1.21\GeV$, and $0.27\GeV$ for the
$\overline{\rm PS}$ mass. So the tendency is going in the proper direction.
However, we stress that the goal, namely to avoid the infrared renormalon
ambiguities, is already reached by the static PS mass as well as by the other
threshold masses which have been developed.

\section{What to do with our result then?}
Having mass values for the top quark at hand, we can consider the cross
section of the process $e^+e^-\to t\bar t$ in the near threshold region where
the velocity $v$ of the top quark is small. It is well known that the
conventional perturbative  expansion does not work in the non-relativistic
region because of the presence of the Coulomb singularities at small
velocities $v\to 0$. The terms proportional to $(\alpha_s/v)^n$ appear due to
the instantaneous Coulomb interaction between the top and the antitop quark.
The standard technique for resumming the terms $(\alpha_s/v)^n$ consists in
using the Schr\"odinger equation for the quark-antiquark
potential~\cite{FadinKhoze}
\begin{equation}
(H-E-i\Gamma)G(\vec r,\vec r\,'|E+i\Gamma)=\delta^3(r-r')
\end{equation}
where $H$ is the non-relativistic Hamiltonian of the heavy quark-antiquark
system, to find the Green function. The Green function is then related
to the relative cross section by the optical theorem,
\begin{eqnarray}
R&=&\frac{\sigma(e^+e^-\to t\bar t)}{\sigma(e^+e^-\to \mu^+\mu^-)}
  \ =\nonumber\\
  &=&e^2_QN_c\frac{24\pi}sC(r_0){\rm Im}\left[\left(1-\frac{\vec p\,^2}{3m^2}
  \right)G(r_0,r_0|E+i\Gamma)\right]\Bigg|_{r_0\to 0}
\end{eqnarray}
where $\sigma(e^+e^-\to\mu^+\mu^-)=4\pi\alpha^2/3s$, $e_Q$ is the electric
charge of the top quark, $N_c$ is the number of colours, $\sqrt{s}=2m+E$ is
the total center-of-mass energy of the quark-antiquark system, $m$ is the top
quark pole mass and $\Gamma$ is the top quark width. The unknown short
distance coefficient $C(r_0)$ can be fixed by using a direct QCD calculation
of the vector vertex at NNLO in the so-called intermediate
region~\cite{CzarneckiMelnikov,BenekeSmirnov} and by using the direct matching
procedure suggested in Ref.~\cite{Hoang}.

If we take the PS mass as input parameter instead of the pole mass, the energy
value $E$ has to be changed to $\sqrt s-2m_{\rm PS}(\mu_f)$ instead of
$\sqrt s-2m$. The same holds for the $\overline{\rm PS}$ mass. It is
shown~\cite{HoangTeubner,MelnikovYelkhovsky,Yakovlev} that the Schr\"odinger
equation can be reduced to the equation
\begin{equation}\label{se1}
(H_1-E_1)G_1(r,r'|E_1)=\delta^3(r-r')
\end{equation}
with the energy $E_1=\bar E+\bar E^2/4m$, $\bar E=E+i\Gamma$, and with the
Hamiltonian 
\begin{eqnarray}
H_1&=&\frac{\vec p\,^2}{m}+V_C(r)+\frac{3\bar E}{2m}V_0(r)
  -\left(\frac23+\frac{C_A}{C_F}\right)\frac{V_0^2(r)}{2m},\nonumber\\
V_0(r)&=&-\frac{\alpha_s(\mu)C_F}r,\nonumber\\
V_C(r)&=&V_0(r)\Bigg\{1+\frac{\alpha_s(\mu)C_F}{4\pi}
  \left(2\beta_0\ln(\mu'r)+a_1\right)\,+\\&&\kern-24pt
  +\pfrac{\alpha_s(\mu)}{4\pi}^2\left(\beta_0^2
  \left(4\ln^2(\mu'r)+\frac{\pi^2}3\right)+2(\beta_1+2\beta_0a_1)\ln(\mu'r)
  +a_2\right)\Bigg\}\nonumber
\end{eqnarray}
where $\mu'=\mu e^{\gamma_E}$, $\mu$ is the renormalization scale, and
$\gamma_E$ is Euler's constant. The final expression for the cross section is
then given by
\begin{equation}\label{final}
R=\frac{8\pi}{m^2}C(r_0)
  {\rm Im}\left[\left(1-\frac{5\bar E}{6m}\right)G_1(r_0,r_0|E_1)\right]
  \Bigg|_{r_0\rightarrow 0}
\end{equation}
with $G_1(r_0,r_0|E_1)$ being the solution of Eq.~(\ref{se1}) at $r=r'=r_0$.
For the numerical solution we use the program derived in Ref.~\cite{Yakovlev}.
Note that we do not take into account an initial photon radiation which would
change the shape of the cross section. This can be easily included in the
Monte Carlo simulation.

\begin{figure}[ht]
\centerline{\epsfig{file=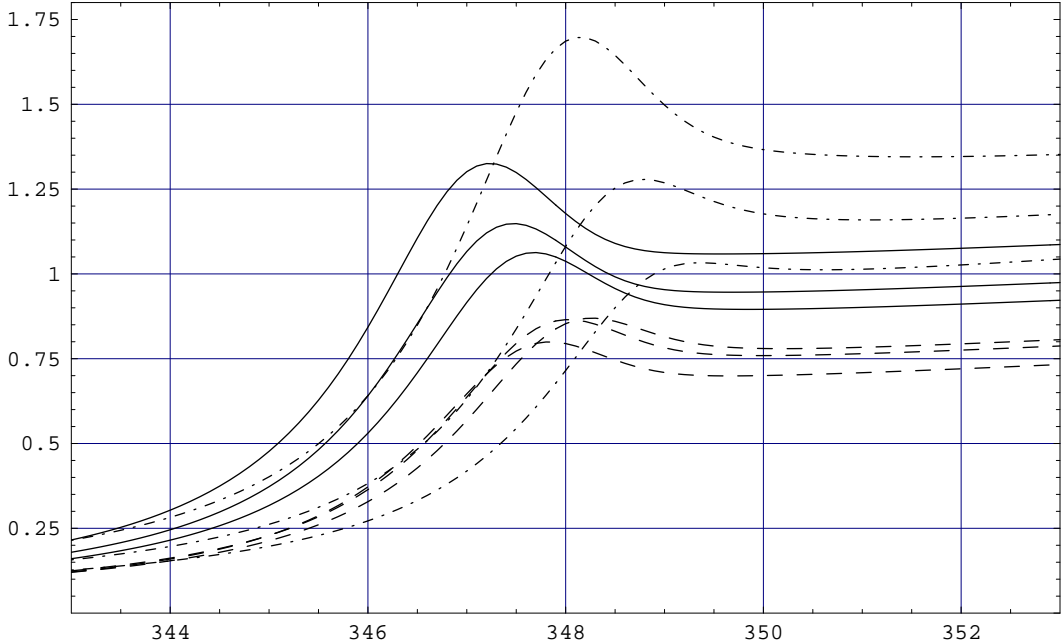,scale=0.85}}
\caption{\label{fig4}The scheme using the pole mass:
shown is the relative cross section $R(e^+e^-\to t\bar t)$ as a function of
the center-of-mass energy in $\GeV$ for the LO (dashed-dotted lines), NLO
(dashed lines), and NNLO (solid lines) approximation. We take the value
$m_t=175.05\GeV$ for the pole mass of the
top quark, $\Gamma_t=1.43\GeV$ for the top quark width,
$\alpha_s(m_Z)=0.119$ and different values
$\mu=15\GeV$, $30\GeV$, and $60\GeV$ for the renormalization scale.}
\end{figure}

Taking the pole mass as input parameter, the top quark cross section at LO,
NLO, and NNLO is shown in Fig.~\ref{fig4} as a function of the center-of-mass
energy. For the top quark pole mass we use $m_t=175.05\GeV$, for the top quark
width $\Gamma_t=1.43\GeV$, and for the QCD coupling constant
$\alpha_s(m_Z)=0.119$~\cite{PDG}. Different values $\mu=15\GeV$, $30\GeV$, and
$60\GeV$ for the renormalization scale are selected because they roughly
correspond to the typical spatial momenta for the top quark. We see that the
NNLO curve modifies the line shape by the amount of $20-30\%$ which is as
large as the NLO correction. It also shifts the position of the $1S$ peak by
approximately $600\MeV$. These large shifts of the peak position are expected
because of the renormalon ambiguity.

\begin{figure}[ht]
\centerline{\epsfig{file=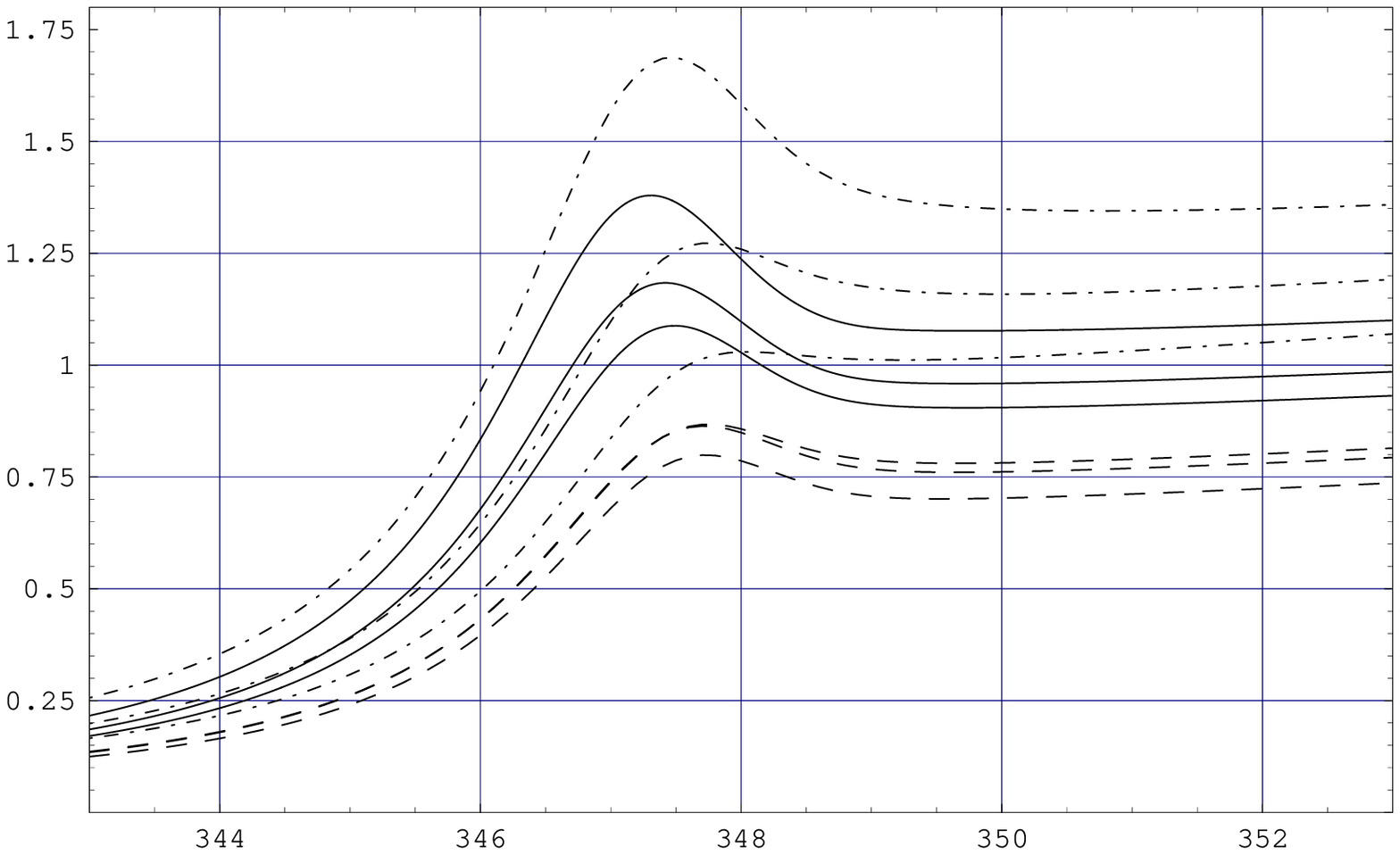,scale=0.85}}
\caption{\label{fig5}The scheme using the PS mass:
shown is the relative cross section $R(e^+e^-\to t\bar t)$ as a function of
the center-of-mass energy in $\GeV$ for the LO (dashed-dotted lines), NLO
(dashed lines), and NNLO (solid lines) approximation. We take the value
$m_{\rm PS}=173.30\GeV$ for the PS mass of the
top quark, $\Gamma_t=1.43\GeV$ for the top quark width,
$\alpha_s(m_Z)=0.119$ and different values
$\mu=15\GeV$, $30\GeV$, and $60\GeV$ for the renormalization scale.}
\end{figure}

If we do the same for the PS mass value $m_{\rm PS}(\mu_f=20\GeV)=173.30\GeV$
which corresponds to the pole mass $m=175.05\GeV$ according to the ``large
$\beta_0$'' accuracy estimate, we obtain a picture like in Fig.~\ref{fig5}
where we observe an improvement in the stability of the position of the first
peak in comparison to the previous analysis as we go from LO to NLO to NNLO.
Actually, for the three values $\mu=15\GeV$, $30\GeV$, and $60\GeV$ we obtain
the maxima of the $1S$ peak for NNLO at $s_{\rm max}=347.32\GeV$, $347.41\GeV$,
and $347.48\GeV$ while the maximal values are given by $R_{\rm max}=1.379$,
$1.184$, and $1.088$, respectively. This demonstrated that a large variation
in the renormalization scale $\mu$ gives rise only to a shift of about
$160\MeV$ for the $1S$ peak position at NNLO while the variation for
$R_{\rm max}$ is still large. The same procedure done for the
$\overline{\rm PS}$ mass shows no observable difference. The differences shown
in Table~\ref{tab1} are too small to make any viewable effect.

\section*{Conclusion and discussion}
We discussed the potential subtracted (PS) mass suggested in Ref.~\cite{Beneke}
as a definition of the quark mass alternative to the pole mass. In contrast to
the pole mass, this mass is not sensitive to the non-perturbative QCD effects.
We have derived recoil corrections to the relation of the pole mass with the
PS mass. In addition, we demonstrated that, if we use the PS or the
$\overline{\rm PS}$ mass in the calculations, the perturbative predictions
for the cross section become much more stable at higher orders of QCD (shifts
are below $0.1\GeV$). This understanding removes one of the obstacles for an
accurate top mass measurement and one can expect that the top quark mass will
be extracted from a threshold scan at NLC with an accuracy of about
$100-200\MeV$.\\[12pt]
{\bf Acknowledgements:}
We are grateful to Ratindranath Akhoury, Ed Yao, Martin Beneke and Andre Hoang
for valuable discussions. O.Y.\ acknowledges support from the US Department of
Energy (DOE). S.G.\ acknowledges a grant given by the DFG, FRG, he also would
like to thank the members of the theory group at the Newman Lab for their
hospitality during his stay.


\begin{thebibliography}{99}
\bibitem{Tevatron}G.~Brooijmans [CDF and D0 Collaborations],\\
  ``Top quark mass measurements at the Tevatron'' [hep-ex/0005030]
\bibitem{Peskin}M.E.~Peskin, ``Physics goals of the linear collider''
  [hep-ph/9910521]
\bibitem{YG}Oleg Yakovlev and Stefan Groote, ``Top quark mass definition
  and $t\bar t$ production near threshold at the NLC'' [hep-ph/0008156]
\bibitem{Review}A.H.~Hoang {\it et al.},
  Eur.~Phys.~J.\ direct {\bf C3} (2000) 1 [hep-ph/0001286]
\bibitem{ChetyrkinSteinhauser}K.G.~Chetyrkin and M.~Steinhauser,\\
  Phys.~Rev.~Lett.\ {\bf 83} (1999) 4001 [hep-ph/9907509];\\
  Nucl.~Phys.\ {\bf B573} (2000) 617 [hep-ph/9911434]
\bibitem{MelnikovRitbergen}Kirill Melnikov and Timo van Ritbergen,\\
  Phys.~Lett.\ {\bf 482 B} (2000) 99 [hep-ph/9912391]
\bibitem{AgliettiLigeti}U.~Aglietti and Z.~Ligeti,
  Phys.~Lett.\ {\bf 364 B} (1995) 75 [hep-ph/9503209]
\bibitem{BrodskyLepageMackenzie}S.J.~Brodsky, G.P.~Lepage and P.B.~Mackanzie,\\
  Phys.~Rev.\ {\bf D28} (1983) 288
\bibitem{BenekeBraun}M.~Beneke and V.M.~Braun,\\
  Phys.~Lett.\ {\bf 348 B} (1995) 513 [hep-ph/9411229]
\bibitem{HoangTeubner}A.H.~Hoang and T.~Teubner,\\
  Phys.~Rev.\ {\bf D60} (1999) 114027 [hep-ph/9904468]
\bibitem{Lepage}W.E.~Caswell and G.P.~Lepage,
  Phys.~Lett.\ {\bf 167 B} (1986) 437;\\
  G.T.~Bodwin, E.~Braaten and G.P.~Lepage,\\
  Phys.~Rev.\ {\bf D51} (1995) 1125 [hep-ph/9407339]
\bibitem{PinedaSoto}A.~Pineda and J.~Soto,
  Phys.~Rev.\ {\bf D59} (1999) 016005 [hep-ph/9805424]
\bibitem{Bigi}I.I.~Bigi, M.A.~Shifman, N.G.~Uraltsev and A.I.~Vainshtein,\\
  Phys.~Rev.\ {\bf D50} (1994) 2234 [hep-ph/9402360]
\bibitem{Beneke}M.~Beneke,
  Phys.~Lett.\ {\bf 434 B} (1998) 115 [hep-ph/9804241]
\bibitem{Gupta} S.N.~Gupta and F.~Radford,\\
  Phys. Rev. {\bf D24} (1981) 2309; {\bf D25} (1982) 3430;\\
  S.N.~Gupta, F.~Radford and W.W.~Repko, Phys.~Rev.\ {\bf D26} (1982) 3305
\bibitem{Duncan}A.~Duncan, Phys.~Rev.\ {\bf D13} (1973) 2866
\bibitem{Fischler}W.~Fischler, Nucl.~Phys.\ {\bf B129} (1977) 157
\bibitem{Peter}M.~Peter, Nucl.~Phys.\ {\bf B501} (1997) 471 [hep-ph/9702245]
\bibitem{Schroder}Y.~Schr\"oder,
  Phys.~Lett.\ {\bf 447 B} (1999) 321 [hep-ph/9812205]
\bibitem{PDG}Particle Data Group (C.~Caso {\it et al.}),   
  Eur.~Phys.~J.\ {\bf C3} (1998) 1
\bibitem{FadinKhoze}V.S.~Fadin and V.A.~Khoze,
  JETP Lett.\ {\bf 46} (1987) 525;\\
  V.S.~Fadin and V.A.~Khoze, Sov.~J.~Nucl.~Phys.\ {\bf 48} (1988) 309
\bibitem{CzarneckiMelnikov}A.~Czarnecki and K.~Melnikov,\\
  Phys.~Rev.~Lett.\ {\bf 80} (1998) 2531 [hep-ph/9712222]
\bibitem{BenekeSmirnov}M.~Beneke and V.A.~Smirnov,\\
  Nucl.~Phys.\ {\bf B522} (1998) 321 [hep-ph/9711391];\\
  M.~Beneke, A.~Signer and V.A.~Smirnov,\\
  Phys.~Rev.~Lett.\ {\bf 80} (1998) 2535 [hep-ph/9712302]
\bibitem{Hoang}A.H.~Hoang,
  Phys.~Rev.\ {\bf D56} (1997) 5851 [hep-ph/9704325];\\
  Phys.~Rev.\ {\bf D57} (1998) 1615 [hep-ph/9702331]
\bibitem{MelnikovYelkhovsky}K.~Melnikov and A.~Yelkhovsky,\\
  Nucl.~Phys.\ {\bf B528} (1998) 59 [hep-ph/9802379]
\bibitem{Yakovlev}O.~Yakovlev,
  Phys.~Lett.\ {\bf 457 B} (1999) 170 [hep-ph/9808463]
\end{thebibliography}
\end{document}